\documentclass[%
 reprint,
 amsmath,amssymb,
 aps,
]{revtex4-1}

\usepackage{graphicx}
\usepackage{dcolumn}
\usepackage{bm}
\usepackage{dblfloatfix}

\begin{document}


\title{Re-measurement of the ${}^{33}$S(\boldmath$\alpha$,p)${}^{36}$Cl cross section for Early solar system enrichment}

\author{Tyler Anderson}
\email{tander15@nd.edu}
\author{Michael Skulski}%
\author{Adam Clark}
\author{Austin Nelson}
\author{Karen Ostdiek}
\author{Philippe Collon}

\affiliation{%
 University of Notre Dame\\
}%

\author{Greg Chmiel$^{1}$, Tom Woodruff$^{1}$, Marc Caffee$^{1,2}$}
\affiliation{
 $^{1}$Department of Physics and Astronomy/PRIME Lab, Purdue University
\\
$^{2}$Department of Earth, Atmospheric, and Planetary Sciences, Purdue University
}%

\date{\today}

\begin{abstract}
Short-lived radionuclides (SLRs) with half-lives less than 100 Myr are known to have existed around the time of the formation of the solar system around 4.5 billion years ago. Understanding the production sources for SLRs is important for improving our understanding of processes taking place just after solar system formation as well as their timescales. Early solar system models rely heavily on calculations from nuclear theory due to a lack of experimental data for the nuclear reactions taking place. In 2013, Bowers et al. measured ${}^{36}$Cl production cross sections via the ${}^{33}$S($\alpha$,p) reaction and reported cross sections that were systematically higher than predicted by Hauser-Feshbach codes. Soon after, a paper by Peter Mohr highlighted the challenges the new data would pose to current nuclear theory if verified. The ${}^{33}$S($\alpha$,p)${}^{36}$Cl reaction was re-measured at 5 energies between 0.78 MeV/A and 1.52 MeV/A, in the same range as measured by Bowers et al., and found systematically lower cross sections than originally reported, with the new results in good agreement with the Hauser-Feshbach code TALYS. Loss of Cl carrier in chemical extraction and errors in determination of reaction energy ranges are both possible explanations for artificially inflated cross sections measured in the previous work.

\end{abstract}

\pacs{Valid PACS appear here}
\maketitle


\section{\label{sec:intro}Introduction}

${}^{36}$Cl is one of many short-lived radionuclides (SLRs) known to be present in the Early solar system (ESS), along with ${}^{26}$Al and ${}^{60}$Fe, all of which are useful as chronometers for astrophysical processes occurring within our Solar system \cite{ess_slrs}. Evidence for the presence of ${}^{36}$Cl ($ t_{1/2} = 0.301$ Myr) in the ESS was found in chondrules and Ca-Al-rich inclusions, specifically from carbonaceous chondrites. Correlations were made between excesses in a ${}^{36}$Cl decay product, ${}^{36}$S, and Cl/S ratios measured in Cl-rich minerals present in meteors \cite{cai1,cai2,cai3,cai4}.  All three of these nuclides, however, have measured abundances above predictions from galactic steady-state enrichment suggesting some additional sources of nucleosynthesis \cite{overabundance}. \\
\indent Possible explanations for these observed overabundances are either injection of stellar nucleosynthesis products from outside the solar system \cite{ess_nuc1,ess_nuc2,ess_nuc3}, or local production via irradiation of gas and dust around the proto-Sun \cite{xwind}. Currently, none of these models can predict the measured abundances for all SLRs simultaneously, due in part to the models' strong reliance on nuclear theory for production cross sections, resulting from a lack of experimental data \cite{ess_slrs,slr_abund}. The uncertainty in commonly-used Hauser-Feshbach calculations is up to a factor of three, and has a significant impact on the predictive power of models \cite{ess_err}.\\
\indent To expand the database for these important reactions, Bowers et al. measured cross sections for ${}^{33}$S($\alpha$,p)${}^{36}$Cl between 0.70 and 2.42 MeV/A \cite{bowers}. The reported data show Hauser-Feshbach codes TALYS and NON-SMOKER systematically under-predicting cross sections across the energy range, regardless of the input parameters. Shortly after publication, a paper by Peter Mohr suggested that, if verified, Bowers' results would pose an extreme challenge to current nuclear theoretical models \cite{mohr}. Therefore, there is a compelling reason to re-measure the reaction, primarily focusing on reviewing the procedure and eliminating any potential sources of error. \\
\indent The details of the experiment are laid out in three parts comprising activations, extraction chemistry, and sample measurement with Accelerator Mass Spectrometry (AMS), and a discussion of the discrepancies between this and the previous work follows. \\

\section{\label{sec:exp}Experimental Procedure}
\subsection{\label{sec:activations}Activations}
\begin{figure}[tbp]
    \includegraphics[width=0.48\textwidth]{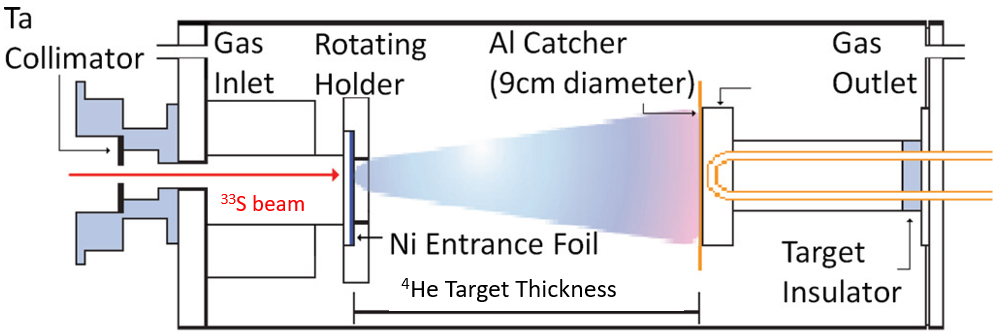}
    \caption{Reproduced from \protect\cite{bowers}, a schematic of the gas cell used for the activations. A 9cm diameter Al foil was used to catch the forward-recoiled ${}^{36}$Cl atoms.}
    \label{figure:gas_cell}
\end{figure}

\begin{figure}[htbp]
    \includegraphics[width=0.48\textwidth]{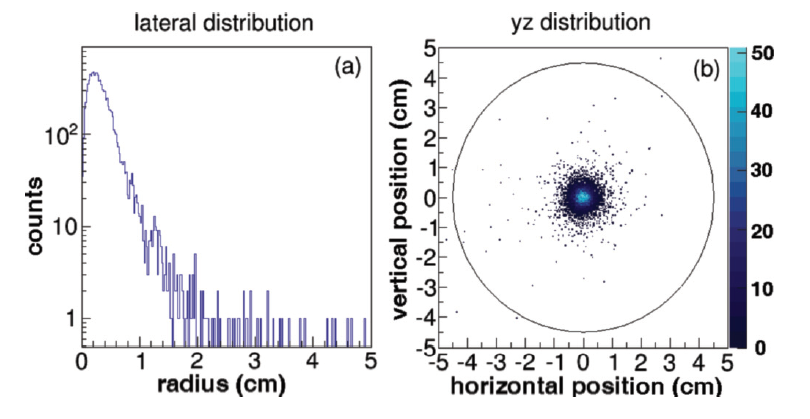}
    \caption{Reproduced from \protect\cite{bowers}, a SRIM simulation of $10^4$ ions of ${}^{33}$S passing through the ${}^{4}$He filled gas cell for the original highest energy sample (104.5 MeV). (a) A histogram of the lateral distribution of ions implanted in the catcher foil, with the beam incident on the gas cell at radius = 0. (b) A 2D histogram showing the distribution of ions across the 9cm diameter cross section of the catcher foil present to the beam. }
    \label{figure:TRIM}
\end{figure}

${}^{36}$Cl was produced in the same way as in \cite{bowers} via ${}^{33}$S($\alpha$,p) through 7 activations ranging in energy from 0.78 to 1.52 MeV/nucleon on target by accelerating a ${}^{33}$S beam on to a ${}^{4}$He filled gas cell, shown in Figure \ref{figure:gas_cell}.
This method has been used successfully before in the study of the ${}^{40}$Ca($\alpha$,$\gamma$)${}^{44}$Ti reaction \cite{act-ams,danthesis} and others. The activations were performed at the Nuclear Science Laboratory at Notre Dame.\\ 
\indent Using an MC-SNICS, a ${}^{33}$S beam was produced from an FeS cathode, accelerated through the FN Tandem Van de Graaff accelerator and tuned in to a gas cell filled with 10 Torr ${}^{4}$He. When tuning, a Ni entrance foil was removed and an insulator was placed at the back of the cell, separating the foil holder from the gas cell to allow the current to be read. During activations, the Ni foil (2.5$\mu$m thick) was returned to the entrance and the back insulator was removed to allow integration of all incident beam current. The Ni foil was contained in a rotating holder to reduce direct heating. After passing through the entrance foil, the ${}^{33}$S beam could react along the length of the gas-filled volume and forward-recoiled ${}^{36}$Cl atoms were stopped in a 0.25mm thick Al foil at the back of the gas cell.\\ 
\indent SRIM was used to calculate energy lost through the Ni foil and ${}^{4}$He gas, shown in Table \ref{table:de/dx}. The energy range for each activation is $E_{low}$ to $E_{high}$, with
$$E_{high} = E_{foil} + FWHM/2,$$
$$E_{low} = E_{gas} - FWHM/2,$$

\noindent where $E_{foil}$ and $E_{gas}$ are the centroid energy after the Ni foil and after the $^{4}$He gas, respectively, and $FWHM$ is the full width at half maximum of the energy distribution.\\
\indent SRIM simulations show that, for all energies, greater than 99$\%$ of recoils are caught in the 9cm diameter circular cross section of the foil (Figure \ref{figure:TRIM}). Activations were performed at 5 different energies, comprising 7 different samples, details about which are listed in Table \ref{table:act}. Samples 2 and 3 each had paired activations of the same energy, denoted 'a' and 'b', but were produced with different activation lengths to produce differing amounts of ${}^{36}$Cl to serve as a check on potential Cl losses in chemical extraction. \\
\begin{table}[htbp]
\centering
\begin{tabular}{|c|c|c|c|c|c|c|c|} 
 \hline
 Sample & $E_{i}$ & $E_{foil}$ & $E_{gas}$ & $FWHM$ & $E_{high}$ & $E_{low}$ & $\Delta E$ \\ 
 \hline\hline
 1 & 56 & 27.9 & 26.1 & 1.0 & 28.4 & 25.6 & 2.8 \\ 
 2a & 63 & 35.3 & 33.5 & 1.1 & 35.8 & 33.0 & 2.8 \\
 2b & 63 & 35.3 & 33.5 & 1.1 & 35.8 & 33.0 & 2.8 \\
 3a & 67.5 & 40.0 & 38.4 & 1.1 & 40.5 & 37.9 & 2.6 \\
 3b & 67.5 & 40.0 & 38.4 & 1.1 & 40.5 & 37.9 & 2.6 \\  
 4 & 74.25 & 47.3 & 45.7 & 1.0 & 47.8 & 45.2 & 2.6 \\
 5 & 76.5 & 49.8 & 48.1 & 1.1 & 50.3 & 47.6 & 2.7 \\
 \hline
\end{tabular}
\caption{Information on energy loss of the ${}^{33}$S beam as it passed through the gas cell for each sample. $E_i$ is the incident beam energy before passing through the Ni foil, $E_{foil}$ is the mean energy after the Ni foil, $E_{gas}$ is the mean energy after passing through the ${}^{4}$He gas, and $FWHM$ is the full width at half maximum of the beam energy distribution after the He gas. $E_{high}$ and $E_{low}$ are the high and low bounds in reaction energy, calculated as described in section \protect\ref{sec:activations}. $\Delta E$ is the energy range for each reaction energy. All values listed are measured in MeV.}
\label{table:de/dx}
\end{table}

\begin{table}[!bp]
\centering
\small
\setlength\tabcolsep{2pt}
\begin{tabular}{|c|c|c|c|c|c|} 
 \hline
 Sample & $E_{i}$ (MeV) & Activation & Charge & $V_{term}$ & $I_{avg}$ (nA)\\  
 & & Length (h) & state & (MV) & \\[0.5ex]
 \hline\hline
 1 & 56 & 71.88 & 7+ & 7 & 189.70 \\ 
 2a & 63 & 21.58 & 8+ & 7 & 153.73 \\
 2b & 63 & 14.83 & 8+ & 7 & 123.55 \\
 3a & 67.5 & 21.55 & 8+ & 7.5 & 103.12 \\
 3b & 67.5 & 15.26 & 8+ & 7.5 & 183.71 \\ 
 4 & 74.25 & 4.72 & 8+ & 8.25 & 229.57\\
 5 & 76.5 & 6.03 & 8+ & 8.5 & 81.94 \\
 \hline
\end{tabular}
\caption{Activation parameters for all samples. $E_i$ is the energy of the beam before the Ni foil, and $V_{term}$ is terminal voltage on the accelerator.}
\label{table:act}
\end{table}

\subsection{\label{sec:chem}Chemistry}

The initial chemistry performed by Bowers was geared toward bulk rock dissolution, which presented issues when dissolving the Al catcher foils. As a result, the procedure was significantly changed and the differences are explained further in section \ref{sec:disc}.\\
\indent First, each of the Al foils, having an average mass of 5.678g, were cut in to smaller pieces with a pair of scissors washed with ethanol and DI water between samples. The foil pieces were then added to separate bottles along with a stable Cl carrier (1.001 mg/g Cl) enriched in ${}^{35}$Cl, with details listed in Table \ref{table:chem}. To begin the dissolution, 40g of H$_2$O was added as a buffer, then 12g HF (49\%) were added to begin the reaction. Another 33g of HF were added in 3g increments, for a total of 45g, allowing the reaction to slow between each addition. After the HF was fully added, 50ml of DI H$_2$O was added to dilute the aluminum fluoride resulting from the dissolution of the foils and prevent formation of an AlF$_2$ gel which would trap some sample. Precipitation as AgCl was performed with addition of AgNO$_3$. The precipitate was compacted in a centrifuge and the excess liquid above the AgCl pellet was decanted. The AgCl was washed several times by breaking the pellet, rinsing with DI water, centrifuging to compact again and repeating. The AgCl was then left to dry in an oven at 80${}^{\circ}$ C overnight. \\
\indent The AgNO$_3$ was added in excess such that every Cl atom  could pair with an Ag atom. Comparing the masses of added Cl to the final AgCl powder resulted in a measured sample collection efficiency of around 97.5\% overall. The majority of the loss was from sample 3b, which was the only sample to show an unidentified green compound after dissolution.

\begin{table}[htbp]
\centering
\begin{tabular}{|c|c|c|} 
 \hline
 Sample & $M_{carrier}$ (g) & $N_{Cl}$ ($10^{20}$ atoms) \\ [0.5ex] 
 \hline\hline
 1 & 8.9308 & 1.51(2)  \\ 
 2a & 8.9806 & 1.52(2)  \\
 2b & 8.9697 & 1.52(2)  \\
 3a & 8.9950 & 1.53(2)  \\
 3b & 8.9724 & 1.53(2) \\ 
 4 & 8.9687 & 1.53(2) \\
 5 & 8.9880 & 1.53(2)\\
 Chem Blank & 4.4960 & 0.765(7)\\ 
 \hline
\end{tabular}
\caption{The amount of stable Cl carrier added (1.001mg/g Cl), and the equivalent number of Cl atoms for each sample.}
\label{table:chem}
\end{table}

\subsection{\label{sec:ams}Accelerator Mass Spectrometry}
The AMS measurements were initially planned to be performed at the NSL using the FN Tandem, but by the time the samples were prepared, the accelerator was shut down in preparation for an upgrade to its low energy injection system. Instead, the AMS measurements were performed at Purdue's Rare Isotope Measurement (PRIME) Laboratory. \\
\indent For each sample, the extracted AgCl powder was pressed on to the surface of AgBr in cathodes warmed on a hot plate to drive away moisture. The AgBr was used for its low sulfur content as isobaric suppression of ${}^{36}$S. The Cl was extracted from the cathode and accelerated using an FN Tandem, identical to the model used in the NSL, to produce a ${}^{36}$Cl beam at 84.3 MeV. After acceleration, the beam was analyzed by a high energy analyzing magnet and then passed through a series of 3 Wien filters for isotopic separation. Isobaric separation of ${}^{36}$Cl from ${}^{36}$S was performed in a 135 degree magnet, filled with 4 Torr of N$_2$ gas. Located immediately after the gas-filled magnet was an ionization chamber filled with 85 Torr P-10 for particle identification. \\

\section{\label{sec:results}Results}
\begin{figure}[htbp]
    \includegraphics[width=0.48\textwidth]{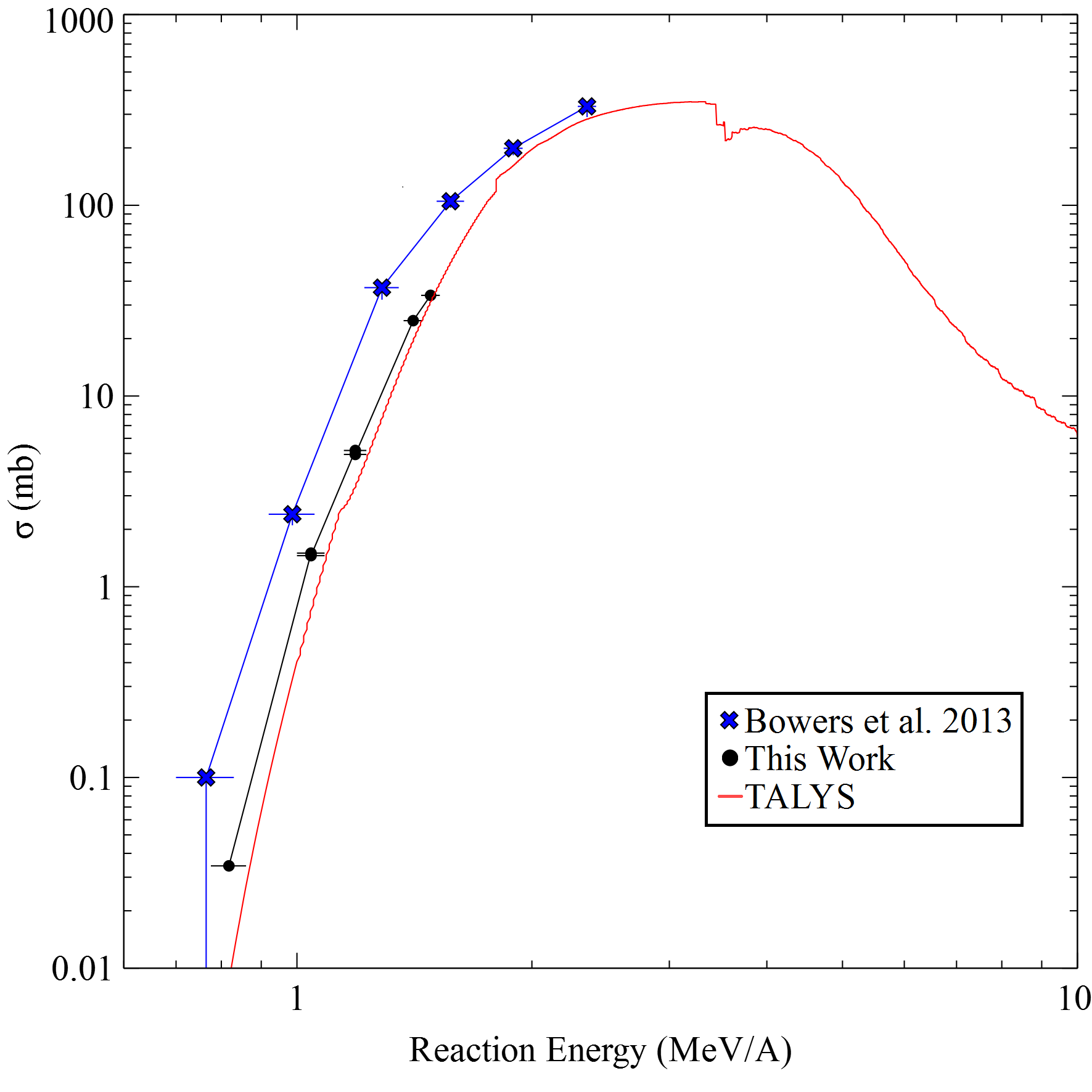}
    \caption{A comparison of the current work's measurements (black), those of Bowers et al. (blue), and cross sections calculated by TALYS (red). Note the two overlapping pairs of points above 1 MeV/A which show the agreement between samples 2a and 2b, and 3a and 3b.}
    \label{figure:results}
\end{figure}

The cross sections were calculated using
$$<\sigma> = \frac{N_{36Cl}}{N_{33} \times N_{T}},$$ where $N_{33}$ is the total number of incident ${}^{33}$S ions for an activation, $N_{T}$ is the areal density of ${}^{4}$He target atoms, and $N_{36Cl}$ is the number of ${}^{36}$Cl atoms calculated in the sample. $N_T$ is given by 
$$N_T=\rho_{atm}\frac{P}{P_{atm}}\frac{N_A}{M_{He}}d,$$ 
where $N_T$ is in units of target nuclei/cm$^2$, $\rho_{atm}$ (=166.3 g/$m^3$) is the density of ${}^4$He at atmospheric pressure, and $P$ and $P_{atm}$ are gas cell and atmospheric pressures, respectively. $N_A$ is Avogadro's constant, $M_{He}$ is the atomic mass of helium (=4.0026 g/mol) and $d$ (=24cm) is the distance between the Ni entrance foil and Al catcher foil in the gas cell.\\ 
\indent The results of the AMS measurement, the reaction energy ranges, and the calculated cross sections for each sample are listed in Table \ref{table:results}. The uncertainties in the measurements are listed in Table \ref{table:error}. These cross sections are plotted with those reported by Bowers et al. and TALYS in Figure \ref{figure:results}. Our new results lie just above the theoretical data from TALYS, with higher energy samples approaching TALYS increasingly closer. This seems to be inconsistent with the generally observed trend of Hauser-Feshbach codes slightly over-predicting cross sections \cite{mohr}.

\begin{table}[htbp]
\centering
\begin{tabular}{|c|c|c|c|c|} 
 \hline
 Sample & $E_{low} - E_{high}$ & ${}^{36}Cl/Cl$ & $N_{36Cl}$ & $< \sigma >$\\ 
 & (MeV/A)& $(10^{-15})$ & $(10^{8})$ & (mb)\\ [0.5ex] 
 \hline\hline
 1 & 0.78 - 0.86 & 77.39 & 0.117 & 0.03(2)\\ 
 2a & 1.00 - 1.08 & 812.59 & 1.24 & 1.36(7) \\
 2b & 1.00 - 1.08 & 485.19 & 0.737 & 1.40(7) \\
 3a & 1.15 - 1.23 & 1648.49 & 2.52 & 4.8(2) \\
 3b & 1.15 - 1.23 & 1988.18 & 3.04 & 4.6(2) \\
 4 & 1.37 - 1.45 & 3861.43 & 5.91 & 23(1)\\
 5 & 1.44 - 1.52 & 2387.46 & 3.65 & 31(2) \\
 Chem Blank & - & 4.11 & - & - \\
 \hline
\end{tabular}
\caption{Shown for each sample are, (1) the reaction energy ranges as determined as described in Section \protect\ref{sec:activations}, (2) the ${}^{36}$Cl/Cl concentrations measured by AMS, (3) the corresponding number of ${}^{36}$Cl atoms deduced for the activations, and (4) the calculated integrated cross section.}
\label{table:results}
\end{table}

\begin{table}[htbp]
\centering
Measurement error budget
\begin{tabular}{|c|c|} 
\hline

 Incident ${}^{33}$S ions ($N_{33}$) & 2\% \\ 
 Stable Cl carrier atoms ($N_{Cl}$) & 1\%   \\
 ${}^4$He target density & 2\%   \\
 ${}^{36}$Cl/Cl  & 3-5\%  \\

 \hline
\end{tabular}
\caption{A summary of uncertainties used for the different measurements.}
\label{table:error}
\end{table}
\section{\label{sec:disc}Discussion}

In the process of repeating the measurements of Bowers et al., there were two main discrepancies discovered in the originally performed procedure that could have led to overly inflated cross sections. The discrepancies involved the chemistry and the determination of each reaction's energy range. \\
\indent The original chemistry performed in Bowers et al. involved dissolution of the Al catcher foils in both HF and HNO$_3$. Repetition of these steps resulted in highly exothermic reactions as well as production of HCl gas, signifying losses of Cl and the sample being compromised. In addition, an AlF$_2$ gel formed in some samples, trapping AgCl away from extraction, further compromising the sample. With the help of PRIME Lab, the chemistry was adjusted to minimize Cl losses to HCl and prevent formation of the AlF$_2$ gel.\\ 
\indent To test the new chemistry, pairs of activations were performed at the same energy, but with different amounts of ${}^{36}$Cl produced (samples 2a, 2b and 3a, 3b). If no Cl was lost during dissolution of the Al foils, then both samples in the pair should result in the same final integrated cross section. The AMS results give confidence that there were no Cl losses in the new chemistry, as the paired measurements agree within error.\\
\indent Second, the previous work's energy loss data was a nearly constant 2 MeV lower than values predicted by SRIM for all energies reported. Re-analysis of the data produced the same results. SRIM, however, reports a deviation between theory and experiment for stopping sulfur ions in nickel of less than 2\% for most of the energy range, and up to 5\% for the lowest energy sample. \cite{fors,diwa,srim}. Given the strong agreement between SRIM and experimental measurements for these energies, all energy loss determination for this work was performed with SRIM. \\
\indent Simulations of 100,000 ions in SRIM were performed for each energy from 0 degrees to a maximum of 6.5 degrees, as defined by the geometry of the activation cell. A Gaussian shape was assumed for the distribution of beam trajectories, with the center at 0 degrees and 6.5 degrees representing $3\sigma$. Beam energies and FWHMs were weighted proportionally by their respective angle's place in the distribution. \\
\indent The data from SRIM showed that the FWHM of beam energies did not differ significantly when calculated immediately after the Ni entrance foil and after both the Ni foil and He gas, supporting the assumption made by Bowers et al. Given this, the FWHM as calculated after passing through both the Ni foil and $^{4}$He gas was used in determining both $E_{high}$ and $E_{low}$. Applying these energy loss data in place of the original measurements by Bowers et al. as a 'correction' shifts all data points except S3 and S4 such that their reaction energy ranges overlap with TALYS. \\

\section*{Acknowledgements}
Thanks are extended to PRIME Lab for all of their guidance in developing the chemistry and for measuring our samples at their lab. PRIME Lab personnel acknowledge support from NSF EAR-0844151. This work is supported by National Science Foundation Grant No. NSF PHY-1419765. 

\bibliographystyle{unsrt}
\bibliography{refs}

\end{document}